\documentclass[format=sigconf,letterpaper]{acmart}
\usepackage{multirow,multicol}
\usepackage{color,tabularx, colortbl,amsmath,bm,arydshln}
\usepackage{enumitem,kantlipsum,arydshln}
\usepackage{booktabs,url,tcolorbox,mathtools} \usepackage{newtxmath}

\acmConference{Submitted to CIKM 2021}{November 1--5, 2021}{Virtual event, Australia}
\acmBooktitle{Proceedings of the 30th ACM International Conference on Information and Knowledge Management}
\copyrightyear{2021}
\acmYear{2021}
\setcopyright{acmlicensed}
\acmPrice{XX.XX}
\acmDOI{XX.XXXX/XXXXXXX.XXXXXXX}
\acmISBN{XXX-X-XXXX-XXXX-X/XX/XX}

\definecolor{Gray}{gray}{0.9}

\begin{document}

\title{Predicting Efficiency/Effectiveness Trade-offs for \\ Dense vs. Sparse Retrieval Strategy Selection}

\author{ Negar Arabzadeh }
\affiliation{%
   \institution{University of Waterloo}}
   \email{narabzad@uwaterloo.ca}

\author{Xinyi Yan }
\affiliation{%
   \institution{University of Waterloo}}
   \email{xinyi.yan@uwaterloo.ca	}

\author{ Charles L. A. Clarke }
\affiliation{%
   \institution{University of Waterloo}}
   \email{charles.clarke@uwaterloo.ca}

\begin{abstract}
Over the last few years, contextualized pre-trained transformer models such as BERT have provided substantial improvements on information retrieval tasks by fine-tuning dense low-dimensional contextualized representations of queries and documents in embedding space. On the other hands, traditional sparse retrieval methods such as BM25 rely on high-dimensional, sparse, bag-of-words query representations to retrieve documents. 
While the dense retrievers enjoy substantial retrieval effectiveness improvements compared to  sparse retrievers, they are computationally intensive, requiring substantial GPU resources, and dense retrievers are known to be more expensive from both time and resource perspectives. In addition, sparse retrievers have been shown to retrieve complementary information with respect to dense retrievers, leading to proposals for hybrid retrievers. These hybrid retrievers leverage low-cost, exact-matching based sparse retrievers along with dense retrievers to bridge the semantic gaps between query and documents. In this work, we address this trade-off between the cost and utility of sparse vs dense retrievers by proposing a classifier to select a suitable retrieval strategy (i.e., sparse vs. dense vs. hybrid) for individual queries. Leveraging sparse retrievers for queries which can be answered with sparse retrievers decreases the number of calls to GPUs.  Consequently, while utility is maintained, query latency decreases.  Although we use less computational resources and spend less time, we still achieve improved performance. Our classifier can select between sparse and dense retrieval strategies based on the query alone. We conduct experiments on the MS MARCO passage dataset demonstrating an improved range of efficiency/effectiveness trade-offs between purely sparse, purely dense or hybrid retrieval strategies, allowing an appropriate strategy to be selected based on a target latency and resource budget.

\end{abstract}

\ccsdesc[500]{Information systems~Information retrieval}
\ccsdesc[500]{Information systems~Evaluation of retrieval results}
\ccsdesc[500]{Information systems~Retrieval effectiveness}
\ccsdesc[500]{Information systems~Retrieval efficiency}
\ccsdesc[500]{Information systems~Information retrieval query processing}

\keywords{Sparse Retriever,Dense Retriever,Query Latency,Retrieval Efficiency}

\maketitle
\vspace{-1em}
\section{Introduction}
Information Retrieval (IR) systems have been revolutionized by neural ranking methods. A few years ago, a state-of-the-art ranking stack would have employed a first stage, sparse retriever, such as BM25, which relies on exact term matches to retrieve an initial pool of items for re-ranking by later stages \cite{lafferty2001document,jones2000probabilistic,schutze2008introduction}. Inverted indices allow the entire corpus to be scanned efficiently, with low latency. First stage retrievers of this type have remained standard for IR since the earliest commercial and academic systems.

Recent advances provide an alternative to first-stage retrievers based on the classic inverted-index. Contextualized pre-trained transformer models, such as BERT \cite{devlin2018bert}, enable a new type of first-stage retriever that maps query and documents into low-dimensional dense vectors and exploit maximum inner product search methods to find a document vectors with the greatest similarity to the query vector~\cite{xiong2020approximate,karpukhin2020dense,zhan2020repbert,gao2020complementing}. While these retrievers enjoy substantial retrieval effectiveness improvements compared to  traditional approaches, they still suffer from drawbacks, including increased GPU resources and potentially higher latency. Although on average these dense retrievers outperform the traditional sparse retrievers on retrieval effectiveness, there remain queries for which the sparse retrievers provide superior performance. There is no single retrieval approach that can definitely answer all the queries effectively and efficiently, and selecting a suitable retrieval strategy is a crucial concern for maintaining the trade-off between a system’s effectiveness and efficiency.
The traditional initial retrievers such as BM25 \cite{jones2000probabilistic} have a long history in Information Retrieval. In a modern ranking stack, BM25 is normally paired with a WAND query processing strategy to maximize the efficiency of first-stage retrieval \cite{broder2003efficient}. These first-stage rankers can be called “sparse retrievers”, since they mostly utilize sparse, high-dimensional, bag-of-word representations in their scoring functions to rank the documents for a given query. 
Since sparse retrievers operate through exact-matching between query and document terms, they do not suffice when there is a semantic gap between the query and the corpus language~--~the vocabulary mismatch problem. However, they are  extremely cost-effective and scale easily, even on inexpensive hardware.
Most sparse ranking methods for first-stage sparse retrievers typically employ BM25 as the first stage in a multi-stage ranking stack~\cite{nogueira2019multi,han2020learning,nogueira2019passage,qiao2019understanding,hofstatter2019effect}. Later stages of the ranking stage use higher-cost neural-based rerankers to rerank the documents retrieved by the initial low-cost sparse retriever. 
While this combination of an initial sparse retrieval step, followed by multiple reranking steps, has shown excellent performance, its performance can be further improved by replacing the initial retrieval step by a dense retriever. These dense retrievers use contextualized pre-trained transformer models to map query and documents into an embedding space~\cite{xiong2020approximate,karpukhin2020dense,zhan2020repbert,gao2020complementing,khattab2020colbert}. The association (e.g., dot-product) between the query and documents in this dense low-dimensional embedding space provides a relevance score. The improvement made by replacing the sparse retriever with dense retriever has been shown on many core retrieval tasks and leaderboards such as MS MARCO. MS MARCO is a large-scale collection with a focus on enabling the applications of deep learning methods for information retrieval. Providing a relatively large volume of training data makes MS MARCO a suitable testbed for comparing SOTA retrieval methods. 

At the time of writing,  the top run on the MS MARCO passage retrieval leaderboard is RocketQA \cite{ding2020rocketqa}, a dense retriever which uses a dual encoder as well as a cross encoder architecture in order to learn dense representations of queries and passages. Similarly, the top performing model on the document retrieval leaderboard is ANCE \cite{xiong2020approximate},  a learning mechanism that learns the query and document representation in a siamese network and uses dot product to rank the documents. Documents can be encoded offline to save time, whereas finding the most similar document to the query is a time-consuming stage in dense retrievers that should be done on the fly when query arrives. Recently, nearest neighbor search algorithms such as FAISS support these dense retrievers, scanning millions of candidates in milliseconds \cite{johnson2019billion}. But even with the most efficient nearest-neighbour algorithm, the latency of dense retrievers may still not be comparable to sparse retrievers. In addition, the need for substantial GPU resources makes dense retrievers expensive to run, and may limit their use in high-volume, real-world applications, where resource limitations are always a crucial concern. Therefore, relative to sparse retrievers, dense retrievers may incur greater costs and impact latency. Also, dense retrievers are not able to produce token-level matching signals which may be critical for named entities and other terms for which an exact match is required. 

Some researchers have proposed “Hybrid Retrievers”, which combine sparse and dense retrievers in the hope of gaining some of the advantages of both methods. Some recent approaches leverage sparse retriever lexical information in the training process. For instance \citet{gao2020complementing} train a dense retriever to supplement a sparse retriever by semantic level information. Other researchers have combined the sparse and dense retriever ranking by interpolating the relevance score from each retrieve \cite{lin2020distilling,luan2020sparse,gao2020complementing,kuzi2020leveraging}. However, to the best of our knowledge, there is no work on selecting the appropriate retrieval strategy for individual queries.

In this paper, we investigate strategies for selecting a first-stage retrieval strategy based on the query. Selecting an appropriate strategy provides a trade-off between efficiency and effectiveness. We select from three possible strategies: 1) sparse only, 2) dense only, and 3) hybrid, where both the dense and sparse retrievers are run and their results merged into a single pool for reranking. By selecting the appropriate strategy on a per-query basis, we can leverage the benefits of both sparse and dense retrievers. For some queries, the use of a dense retriever might avoid vocabulary mis-match problems. For other queries, the use of a sparse retriever might provide necessary exact term matches. For some queries, both may be desirable. When a query can be answered with a lower-cost, lower latency, sparse retriever alone, costs may be reduced by avoiding the use of a dense retrieval method. We utilize contextualized pre-trained embedding representation of queries to train the classifier in a cross-encoder architecture with the goal of selecting the retrievers that can answer the query the best. Further, we run experiments on MS MARCO passage collection to train two classifiers i.e., select `sparse vs.\ dense' or `sparse vs.\ hybrid' on per query-basis. Our experiments provides a setting where you can select the best retrieval strategy based on query latency and resource constraints. 

\section{Method}
When selecting the retrieval strategy, we aim to capitalize on the efficiency and parsimony of the sparse retriever, preferring it over the dense retriever when possible. Nonetheless, if our classifier predicts relatively poor performance for the sparse retriever on query $q$, we would prefer to exploit the more powerful and resource hungry dense retrieval strategy. For some queries, we also consider a hybrid approach, by running both retrievers and combining the pools of retrieved items from sparse and dense retrievers into a single pool for reranking. Under this hybrid strategy, we always run the sparse retriever, and then decide to run the dense on the basis of the results.
We investigate the effect of this retrieval strategy selection process on the MS MARCO passage collection\cite{nguyen2016ms}\footnote{\url{ https://microsoft.github.io/msmarco/
}}. This collection consists of 8.8 million passages accompanied by more than 500k pairs of query and judged relevant passages for training purposes. For each query $q$, there is a  set of relevant judged passages $R_q$ where for over 90\% of queries $|R_q|=1$, i.e., there is a single judged relevant passage per query. In addition, there is set of 6,980 queries for development and validation (``MS MARCO Development Set''). Finally, there is a test set with private relevance judgments that are not available to us. In our experiments, we work with the training and development sets.

As our sparse retriever, we employ BM25 as implemented by the open-source Anserini system from the University of Waterloo \cite{yang2017anserini}, which provides state-of-the-art performance for sparse retrievers. The Anserini\footnote{\url{ https://github.com/castorini/anserini
}} implementation of BM25 has been widely used in top-performing reranking stacks~\cite{nogueira2019multi,hofstatter2020interpretable,hofstatter2020local}.  As our dense retriever, we adopt ANCE, which is currently a state-of-the-art first-stage dense retriever from  the standpoint of both efficiency and effectiveness. ANCE has shown to be more than 100 times faster than other dense retrievers\cite{xiong2020approximate}. In addition, we repeat our experiments with other representative dense retrievers, such as RepBERT~\cite{zhan2020repbert}
and ColBERT~\cite{khattab2020colbert}. For our hybrid retriever, we simply retrieve documents by both sparse (BM25) and dense (ANCE) and merge them into a single pool. In the following, we discuss how we train our proposed classifiers to decide between `sparse vs. dense' or `sparse vs. hybrid'  as the retrieval strategy on a per query basis.

\subsection{Sparse vs. Dense}
\label{sparsevsdense}

In selecting a retrieval strategy, we prefer the lower-cost sparse retriever to a dense retriever. For the training queries, we base labels for our classifier on the position of the first relevant passage in the results from the sparse retriever. For a given training query $q$,  we call the set of relevant passages $R_q$ and the ranked list of top-$K$ retrieved documents with sparse retriever $S_q^K$. Let the first relevant retrieved passage within $S_q^K$  be $F_q$. If $F_q$ appears above a threshold $T$ in $S_q^K$, we label $q$ as ``Sparse Retriever''. If $F_q$ appears below the threshold, or if $S_q^K$ does not contain a relevant passage, we label $q$ as ``Dense Retriever'':
\begin{equation}
    S_q^K = [d_1,d_2,d_3, ... , d_K]
    \label{sqk}
\end{equation}
\begin{equation}
   F_q = \{d_x | x= min\{i|d_i \in S_q^K, d_i\in R_q\}\}
   \label{Fq}
\end{equation}
\begin{equation}
Rank(F_q) =  min\{i|d_i \in S_q^K, d_i\in R_q\}
\label{rank}
\end{equation}
\begin{equation}
Strategy=
 \begin{cases}
    Sparse Retriever & \text{if $Rank(F_q) <= T$}\\
    Dense Retriever & \text{otherwise}\\
    \end{cases}    
    \label{schema1}
\end{equation}
For the results in this paper we use a threshold of $T = 50$, but similar results are obtained for other values (100, 150, and 200).

We use these labels to fine-tune BERT. In recent years, BERT has been extensively used with enormous success on a variety of  tasks in Natural Language Processing and Information Retrieval, including first-stage retrieval, reranking, paraphrasing and text similarity~\cite{devlin2018bert,sbert,nogueira2019passage,bert-text-class}.  BERT utilizes a cross-encoder framework, which jointly encodes the inputs and candidates in a single transformer. One of the benefits of having all the inputs in such an architecture is having a high number of  interactions between the input tokens. Although this model suffers from intensive and slow computations when there is a large-scale pool of candidates  at inference time, it  aligns with the needs of our classifier because we do not have a high number of inputs. Each query comprises a small number of tokens. We utilized contextualized pre-trained embedding representation of queries since they consider both semantics and context of the queries. The query tokens followed by special tokens are fed into the cross-encoder network. The model performs full-cross self-attention over the given input and label with the aim of attaining higher accuracy. Further, a linear classification layer with binary cross entropy loss  on top of the first vector produced by the transformer to reduce dimensions and get a scalar value probability for each class. 

\subsection{Sparse vs.\ Hybrid}
\label{sparsevshybrid}
\begin{figure}[t]
  \centering
  \includegraphics[width=0.99\linewidth]{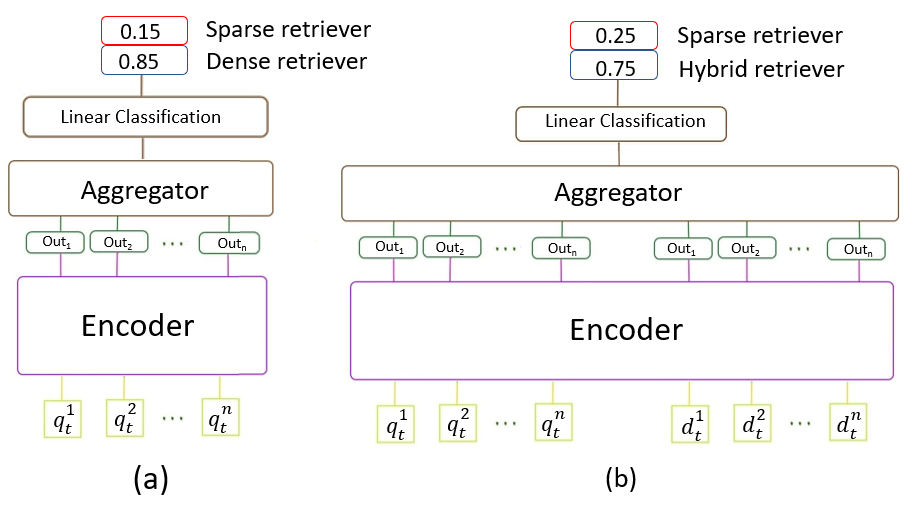}
  \vspace{-1em}
  \caption{ Retrieval category selection classifiers: (a) sparse vs.\ dense, using only the query as input and, (b) sparse vs.\ hybrid, using both query and top-retrieved document as input; $q_t^i$ and $d_t^i$ represent query and document tokens.
}
\vspace{-1em}
    \label{fig:classifiers}
\end{figure}

In the previous section, we describe a classifier to select between a sparse or dense retriever given only the text of a query. In this section, we select between a sparse or hybrid strategy, where the selection in made after an initial sparse retrieval, when the results of the sparse retrieval are available. If hybrid retrieval is selected, dense retrieval is performed as a second step and its results are pooled with those from the sparse retriever. As we did in the previous section, labels for training are based on presence and rank of the top relevant passage $F_q$ returned by the sparse retriever: 
\begin{equation}
Strategy=
 \begin{cases}
    Sparse Retriever & \text{if $Rank(F_q) <= T$}\\
    Hybrid Retriever & \text{otherwise}\\
    \end{cases}    
    \label{schema2}
\end{equation}
In addition to the text of the query, the text of the top passage returned by the sparse retriever is provided as input to the classifier. We compare the architectures of the two classifiers, i.e., sparse vs.\ dense and sparse vs.\ hybrid in Figure~\ref{fig:classifiers}.


\section{Experiments}
\subsection{Experimental Setup}

To train the classifiers, we follow the labelling scheme described in Section~\ref{sparsevsdense} and~\ref{sparsevshybrid}. We assign labels to queries in the training set. For the sparse vs.\ dense classifier we fine-tune BERT base uncased for 1 epoch with batch size of 8. Training on a RTX 2080 GPU took less than 1 hour. For the sparse vs.\ hybrid classifier we fine tune the same pretrained model using the query and the top document retrieved by the sparse retriever.  We feed in the query followed by a separator token, the first retrieved document and the assigned class to fine-tune BERT base uncased with the same experimental setup. We made our code publicly available at \url{https://github.com/Narabzad/Retrieval-Strategy-Selection}. 

\subsection{Results and Findings}

As shown in Figure 1, the classifiers produce probabilities that queries belong to each of the two classes. To specify a strategy, we pick a threshold between 0.0 and 1.0. By picking different thresholds, we can trade off between the efficiency and resource parsimony of the sparse strategy and the improved effectiveness of the dense and hybrid strategies. To measure effectiveness, we report recall@1000, but the overall results are similar at other recall levels. We use recall because the output of this first-stage retrieval will be re-ranked by additional stages. The goal of the first stage is build a pool containing relevant passages, while the goal of later stages is to place these relevant passages at the top of the ranking.

\begin{figure}[t]
  \centering
  \label{fig1}
  \vspace{-1em}
  \includegraphics[width=0.94\linewidth]{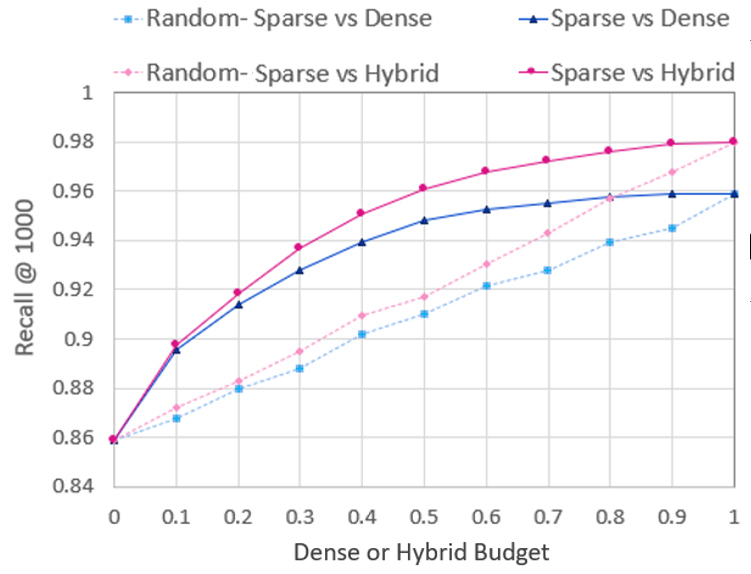}
    \vspace{-1em}
  \caption{ Performance of retrieval strategy selection. The x-axis shows \% of queries retrieved by dense retrievers. 
}
  \label{fig:pref}
    \vspace{-1em}
\end{figure}

Figure~\ref{fig:pref} shows the trade-off between sparse vs.\ dense retrievers. As the threshold is varied from 0.0 to 1.0, the dense retrieval strategy is selected for a larger and larger faction of the queries. The x-axis shows fraction of queries that are assigned the dense retriever across all 6980 queries in MS MARCO development set. We can view this fraction as a ``budget'' allocated to the dense retrieval strategy, which we allocate to improve effectiveness. For a budget of 0, the sparse retriever is always used. For a budget of 1, the alternative (dense or hybrid) retriever is always used. At the lower left, where the two blue curves intersect, the dense retriever budget is 0, meaning all the queries were executed with a sparse retriever. The blue point at the upper right, where the blue curves intersection again shows the performance at a dense retriever budget of 1, which means all the queries were executed with a dense retriever. As a baseline, we randomly assign a retriever at the rate given by the budget, shown by the dotted lines. Both classifiers substantially outperform this baseline. For example, at dense retriever budget of 0.5, when 50\% of queries (3490 queries) employ the dense retriever and the rest of the queries (3490 queries) employ the sparse retriever (blue lines) the random retrieval strategy selector obtains a recall of 0.91 while our classifier obtains a recall of 0.95, a recall improvement of 0.04.

The performance of the second classifier, sparse vs.\ hybrid, is shown in Figure~\ref{fig:pref} with solid pink lines. The lower left point shows the performance of all 6980 queries in the MS MARCO development set when executing all of them with the sparse retriever and the upper right pink point illustrate the performance when executing all of them by the hybrid retriever, i.e., both the sparse and dense retrievers.  We also show the performance of randomly assigning a fraction of the queries to the hybrid retriever as the dashed pink line. For instance, at the hybrid budget of 50\% where we can only execute 50\% of the queries with the hybrid strategy, the random baseline obtained a recall of 0.91 while our the classifier obtained a recall of 0.96, a recall improvement of over 0.05. At this 50\% level, the recall of the sparse vs.\ hybrid strategy exceeds the recall of the dense retrieval strategy at the 100\% level, while using the dense retriever for only half the queries.
Figure~\ref{fig:pref} also illustrates the gap between the full dense strategy and the full hybrid strategy. To reach a recall@1000 of 0.98, both retrievers must be used. While the dense retriever is superior to the traditional sparse retrieval, it still misses relevant passages found by the sparse retriever.

\begin{figure}[t]
  \centering
  \label{fig1}
  \vspace{0em}
  \includegraphics[width=0.95\linewidth]{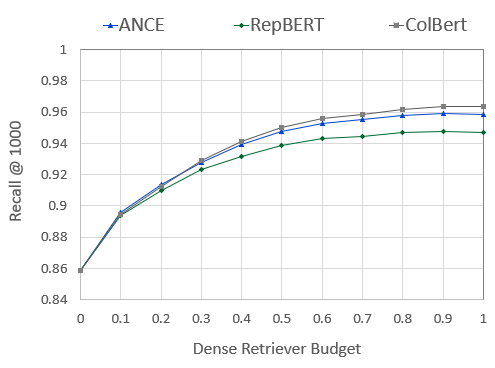}
  \vspace{-1.3em}
  \caption{Performance of retrieval strategy selection with three different dense retrievers.
}
  \label{fig:alt}
\end{figure}


While BM25 is well established as a state-of-the-art sparse retrieval method, there are alternatives we can use as the dense retriever. To test the robustness of our results, we repeat the experiment with other state-of-the-art dense retrievers: Repbert (a representation based dense retriever) \cite{zhan2020repbert} and Colbert (a late-interaction based dense retriever)\cite{khattab2020colbert}. Figure~\ref{fig:alt} shows the result of our sparse vs. dense retrieval strategy with these three dense retrievers. As shown, the type of dense retriever does not impact our conclusions, and our retrieval strategy is robust to the type of dense retriever.

\begin{figure}[t]
  \centering
  \includegraphics[width=0.9\linewidth]{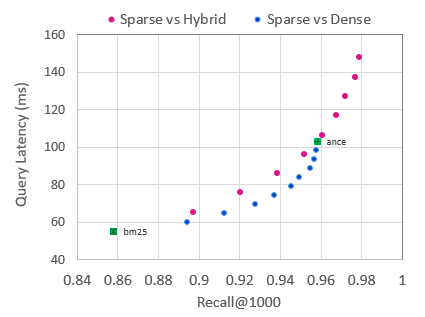}
  \vspace{-1em}
  \caption{Query latency vs.\ recall trade-off for our methods.
}

  \label{fig:latency}
\end{figure}

Finally, we consider the effect of using our retrieval strategy selector in an end-to-end retrieval framework. In \citet{lin2020distilling} query latency time is reported as 55ms for BM25 as the sparse retriever and 103ms ANCE as the dense retriever on the MS MARCO passage collection. As we shift budgets for dense retrieval, from 0\% to 100\%, our proposed method can prioritize queries that need to be retrieved with dense retriever. Figure~\ref{fig:latency} shows the trade-off between latency and effectiveness as we vary the budget. The Pareto frontier shifts between retrieval strategies as latency and recall increase. For example, the sparse vs. dense strategy provides a recall above 0.94 with a latency under 80ms, while the sparse vs. hybrid strategy provides a recall above 0.96 with a latency under 106ms.


\section{Conclusion}

Dense retrievers provide substantial performance improvements when compared to traditional sparse retrievers, such as BM25. However, dense retrievers introduce performance and resource costs not associated with sparse retrievers, including the need for substantial GPU resources. In addition, dense retrievers can still miss relevant items surfaced by sparse retrievers. In view of the efficiency vs.\ effectiveness trade-offs, we present methods for selecting the appropriate retrieval strategy on a per-query basis, allocating the dense retriever to those queries most likely to benefit. Utilizing our proposed retrieval strategy selection leads achieving improved performance even under  computational resources or  time constraint. Using previously reported performance characteristics we illustrate the trade-off between strategies, where the preferred strategy depends on latency. Other performance characteristics will produce a different trade-off.

\newpage
\bibliographystyle{ACM-Reference-Format}
\balance
\bibliography{acmart} 

\end{document}